\documentclass[]{aastex701}

\usepackage{graphicx}	
\usepackage{amsmath}	
\usepackage{acronym}
\usepackage{bm}
\usepackage{makecell}
\usepackage{xcolor}

\acrodef{NS}{Neutron Star}
\acrodef{BH}[BH]{Black Hole}
\acrodef{BBH}[BBH]{Binary Black Hole}
\acrodef{GW}[GW]{gravitational wave}
\acrodef{SNR}{signal-to-noise ratio}

\newcommand{\D}{\mathrm{d}}
\newcommand{\M}{\mathrm{M}}

\begin{document}

\title{Examining the Gap in the Chirp Mass Distribution of Binary Black Holes}

\author[0000-0002-1602-4176]{Vaibhav Tiwari}
\affiliation{Institute of Gravitational Wave Astronomy, \\School of Physics and Astronomy, University of Birmingham, \\ Edgbaston, Birmingham B15 2TT, UK}
\email{vaibhavtewari@gmail.com}



\begin{abstract}

The mass distribution of binary black holes inferred from gravitational-wave measurements is expected to shed light on their formation mechanisms. An emerging structure in the mass distribution indicates the presence of multiple peaks around chirp masses of $8M_\odot$, $14M_\odot$, and $27M_\odot$. In particular, there is a lack of observations between chirp masses of 10 and 12 $M_\odot$. In this letter, we report that observations significantly favour the model that supports suppression of the rate in a narrow chirp mass range over the model that doesn't include suppression, at a confidence greater than 99.5\%. Using another test, which measures the deviation between the inferred chirp mass distributions from the two models, we conservatively estimate a 95\% confidence in the presence of a feature. A lack of evidence has been reported in the component mass distribution around the comparable location. The differing conclusions are due to a unique correlation between the primary~(heavier of the two masses) and the secondary~(lighter of the two masses) masses of binary black holes. This correlation results in increased clustering of measured chirp masses around specific values. However, it remains to be ascertained whether this correlation can be attributed to the weak measurability of features in the component masses.

\end{abstract}



\section{Introduction} \label{sec:intro}

The chirp mass of a compact binary with masses $m_1$ and $m_2$ is given by
\begin{equation}
\mathcal{M} = \frac{\left( m_1 m_2\right)^{3/5}}{\left(m_1 + m_2\right)^{1/5}}.
\end{equation}

\begin{figure*}
    \centering
    \includegraphics[width=0.99\textwidth]{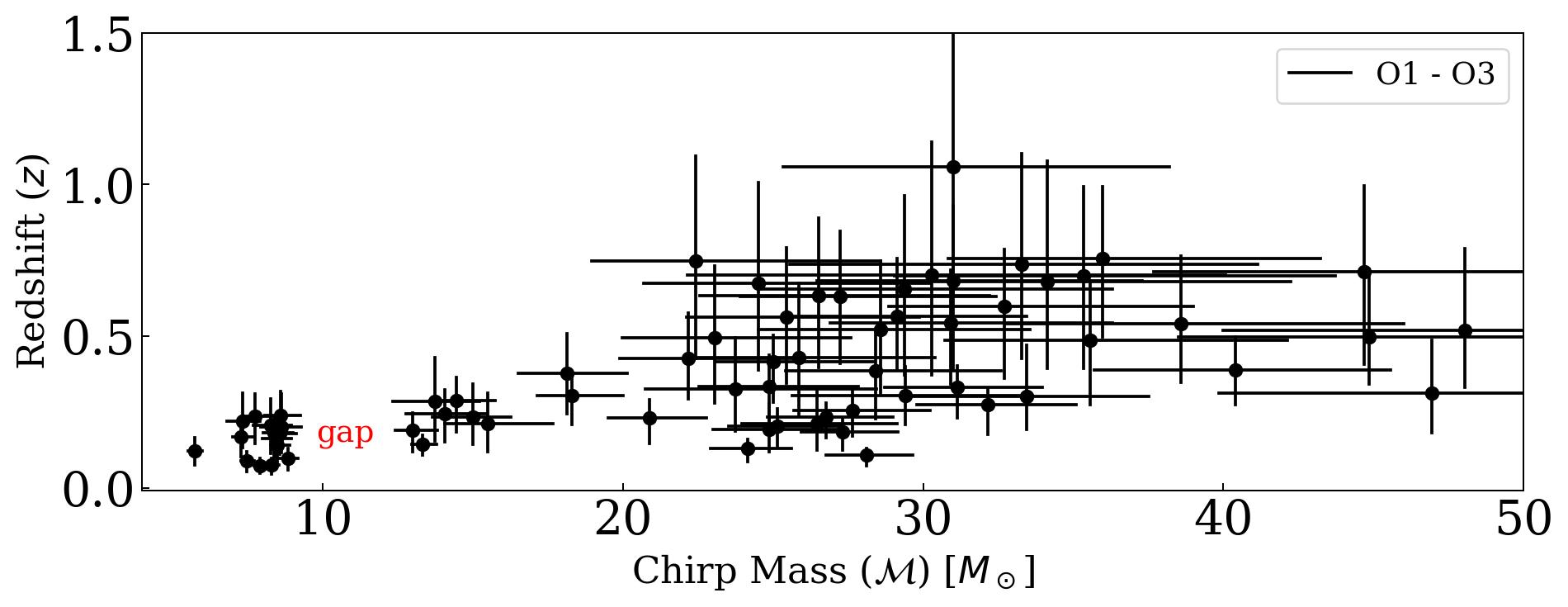}
    \caption{ The measured chirp mass and redshift value of the 69 \ac{BBH} observed with a false alarm rate of once per year or less. Observations cluster around a chirp mass value of 8$M_\odot$ and 14$M_\odot$, leaving a gap in between.}
\label{fig:gap}
\end{figure*}

Chirp mass dominates the phase evolution of \ac{GW} and, therefore, is the most accurately measured parameter~\citep{Cutler:1994ys}. However, it is not expected that channels responsible for forming and merging compact binaries will directly imprint the chirp mass distribution. Any features that do appear will arise indirectly, through the way astrophysical processes shape the distributions of the primary and secondary masses.
The total number of \ac{BBH} observed with a maximum false alarm of one per year after the culmination of LIGO-Virgo-KAGRA~(LVK) collaborations third observation run~(O3) is 69~\citep{2015CQGra..32g4001L, 2015CQGra..32b4001A, 2021PTEP.2021eA101A}. These observations have been observed with a false alarm rate of at least one per year. Figure~\ref{fig:gap} shows these observations' chirp mass/redshift measurements. The chirp mass measurements cluster around specific mass values, resulting in peaks and gaps in the inferred chirp mass distribution~\citep{2021ApJ...913L..19T, 2021ApJ...917...33L, o3b_rnp, 2022ApJ...928..155T, 2024MNRAS.527..298T}. Investigations have also aimed at peaks observed in the primary mass distribution~(e.g. see \cite{2023ApJ...946...16E, 2023MNRAS.524.5844T, 2023ApJ...955..107F, 2023arXiv230401288G, 2023ApJ...957...37R, 2024arXiv240403166R, 2024PhRvL.133e1401L}).

Figure~\ref{fig:gap} shows a lack of observations in the range 10--12$M_\odot$. This range, which we may refer to as \emph{gap}, but could also be a narrow local minima, has been explained as a feature of failed supernova mechanism~\citep{2023A&A...676A..31D} or isolated binary evolution of stripped stars~\citep{2021A&A...645A...5S, 2023ApJ...950L...9S, 2025arXiv251007573W}~(this scenario suggests bimodality in the component mass distribution and the corresponding tri-modality in the chirp mass distribution, with the third intermediate peak in the chirp mass distribution arising from binary black holes with one black hole contribution from each of the two peaks). The unique location of peaks has also been suggested as occurring due to a hierarchical merger scenario when the first peak is formed by the merger of \ac{BH} of stellar origin and successive peaks by the repeated merger of remnants~\citep{2021ApJ...913L..19T, 2022ApJ...928..155T}. Follow-up analyses have fit the prominent peaks in the primary mass distribution using similar proposals~\citep{2025PhRvD.111b3013M}. Such a proposal potentially explains the depleted rate between the two peaks\footnote[1]{However, depending on the rate, intergenerational mergers will fill this gap.}.

Given these unconventional proposals, confidence in the presence of this feature must be well-founded. Presence of a gap in the component mass range 10--15$M_\odot$ was investigated by~\cite{2024ApJ...975..253A}. They reported non-observation of a gap in this mass range. \cite{2025A&A...694A.186G} investigated the presence of \emph{polluting} observations that would fill the gap in the component mass distribution. However, a sharp drop and rise of the inferred chirp mass distribution suggest the presence of a feature and raise the question of how it could leave a gap in the chirp mass distribution, which requires an explanation.

In this article, we focus on the chirp mass range 9.5--20$M_\odot$, at either end of which is the presence of a prominent peak. We estimate the confidence in the presence of a gap in this chirp mass range. This article is laid out as follows. In Section~\ref{sec:method} we describe the method. In Section~\ref{sec:exam}, we discuss our investigations, and in Section~\ref{sec:discuss}, we discuss our findings.

\section{Method} \label{sec:method}
Vamana is a Gaussian mixture model framework to infer the \ac{BBH} population~\citep{2021CQGra..38o5007T}. The components of the mixture model are three-dimensional multivariate normal distributions that model the chirp mass, mass ratio, and aligned spins (components aligned with the orbital angular momentum). Each component also includes a power law to measure the evolution of the merger rate with the redshift. It is a Bayesian analysis; thus, inference is made by drawing many possible distributions calculated from hyperparameter posteriors. We employ three flavours of Vamana to perform our investigation. The first flavour is a mixture model with no constraints on the location of the Gaussian modelling the chirp mass. The second flavour restricts the location of Gaussians and exclusively uses a power-law to infer the distribution in the range 9.5--20$M_\odot$. The second flavour has been set up to avoid inferring a local maximum or minimum in this range. The third flavour is identical to the second, but with additional hyperparameters to suppress density in narrow chirp-mass regions. These flavours are labelled as $\mathbb{M}$, $\mathbb{M}_{pl}$, and $\mathbb{M}_{pl+gap}$ respectively and are described in Appendix~\ref{models}.

We first investigate the presence of a gap by analysing if $\mathbb{M}_{pl}$ has a better or poorer fit with the \ac{GW} observations compared to the two remaining models. A standard approach when comparing two models is to calculate the Bayes factor, which indicates the preferred model. 
\begin{equation}
\mathcal{B} = \frac{p(\mathbf{d}|\mathbb{M}_1)}{p(\mathbf{d}|\mathbb{M}_2)} = \frac{\int p_1(\Lambda)\left[\int \mathcal{L}(\mathbf{d}|\theta)\,p_1(\theta|\Lambda)\, \mathrm{d}\theta\right]\,\mathrm{d}\Lambda}{\int p_2(\Lambda)\left[\int \mathcal{L}(\mathbf{d}|\theta)\,p_2(\theta|\Lambda)\, \mathrm{d}\theta\right]\,\mathrm{d}\Lambda},
\label{eq:BF}
\end{equation}
where $\theta$ are the modeled parameters, and $\mathbf{d}$ is the data. $\Lambda$ and $p(\Lambda)$ are the model hyperparameters and their priors respectively. $\mathcal{L}$ is the likelihood of observing data $\mathbf{d}$ if gravitational waves were emitted by a binary with parameters $\theta$. $p(\theta|\Lambda)$ is the functional form of the model. The term inside the square brackets is the conditional marginal likelihood, which is used in sampling the hyperparameter posterior.

The Bayes factor is sensitive to the choice of the hyperparameter prior. For example, for the model $\mathbb{M}_{pl+gap}$, the prior on the extra hyperparameters controlling the gap can be adjusted to increase or decrease the Bayes factor in comparison to model $\mathbb{M}_{pl}$; thereby incorrectly concluding the presence or absence of a gap. We overcome this situation by integrating the numerator and denominator in Equation~\ref{eq:BF} only for hyperparameters which meaningfully contribute to the posterior. We achieve this by first sampling a fixed number of hyperparameter posteriors for a model and choosing the smallest conditional marginal likelihood as the threshold above which integrals in Equation~\ref{eq:BF} are calculated next. Since most of the priors are uniform, we are effectively calculating the Bayes factors corresponding to the optimal priors for the models. However, the estimated quantity is not a Bayes factor in the usual sense.

We use an additional statistic to assess the presence of a gap: a weighted sum of squared deviations between the chirp mass distributions inferred from model $\mathbb{M}_{pl}$ and the other two models.
We quantify the departure between two distributions by defining a quantity
\begin{align}
\chi^2 =& \sum_{i}\; \frac{\left(\mu_1(\mathcal{M}_i) - \mu_2(\mathcal{M}_i)\right)^2}{\sigma_1(\mathcal{M}_i)\,\sigma_2(\mathcal{M}_i)}
\label{eq:chisq}
\end{align}
where $\mu_1$ and $\mu_2$ are the mean density, and $\sigma_1$ and $\sigma_2$ are the standard deviation of the inferred chirp mass from models 1 and 2. The population from the two models is inferred on discrete mass values, $\mathcal{M}_i$. The summation in Equation~\ref{eq:chisq} is done for values between 9.5$M_\odot$ and 20$M_\odot$. The bins are placed at an interval of 0.1$M_\odot$. The magnitude of $\chi^2$ depends on the difference between the mean distributions inferred from the two models and the accuracy with which these means were estimated. 

For both the statistics, we set the second model to be always $\mathbb{M}_{pl}$, which does not infer a feature, and make comparisons with the distribution inferred from model $\mathbb{M}_{pl+gap}$ or $\mathbb{M}$ capable of inferring local maxima or minima. To convert the statistics to p-values, we use a fiducial chirp mass distribution for simulating multiple catalogs after mimicking the observations and measurement process. We choose the mean inferred distribution from model $\mathbb{M}_{pl}$ as the fiducial distribution. This choice is data-informed and ensures that the simulated catalogs have a comparable number of observations to the real catalog in the chirp mass range of 9.5--20 $ M_\odot$. Similarly, we choose data-informed distributions of mass ratio and aligned spin. Using our models, we infer populations from each of the simulated catalog. By following this procedure, we accumulate BF/$\chi^2$ values for an astrophysical chirp mass distribution assumed to be featureless and that best describes the data. Finally, the BF$_\mathrm{obs}$/$\chi^2_{obs}$ of the \ac{GW} observations are compared against values from the simulated catalogs. We have described the creation of simulated catalogs in Appendix~\ref{sims}.

\section{investigation} \label{sec:exam}

We investigate whether the observations can be confidently explained as just another draw from the fiducial distribution, i.e. in reality, the underlying distribution is featureless and the observed chirp mass values are just one among many possible draws from it. In this section, we perform two comparisons. First, we investigate a gap characterised by sharp decay in density in a narrow chirp mass range using models $\mathbb{M}_{pl}$ and $\mathbb{M}_{pl+gap}$. In our second investigation, we relax the shape of the gap and draw our conclusion by comparing models $\mathbb{M}_{pl}$ and $\mathbb{M}$

\subsection{Comparing $\mathbb{M}_{pl}$ and $\mathbb{M}_{gap+pl}$}

\begin{figure*}
    \centering
    \includegraphics[width=\textwidth]{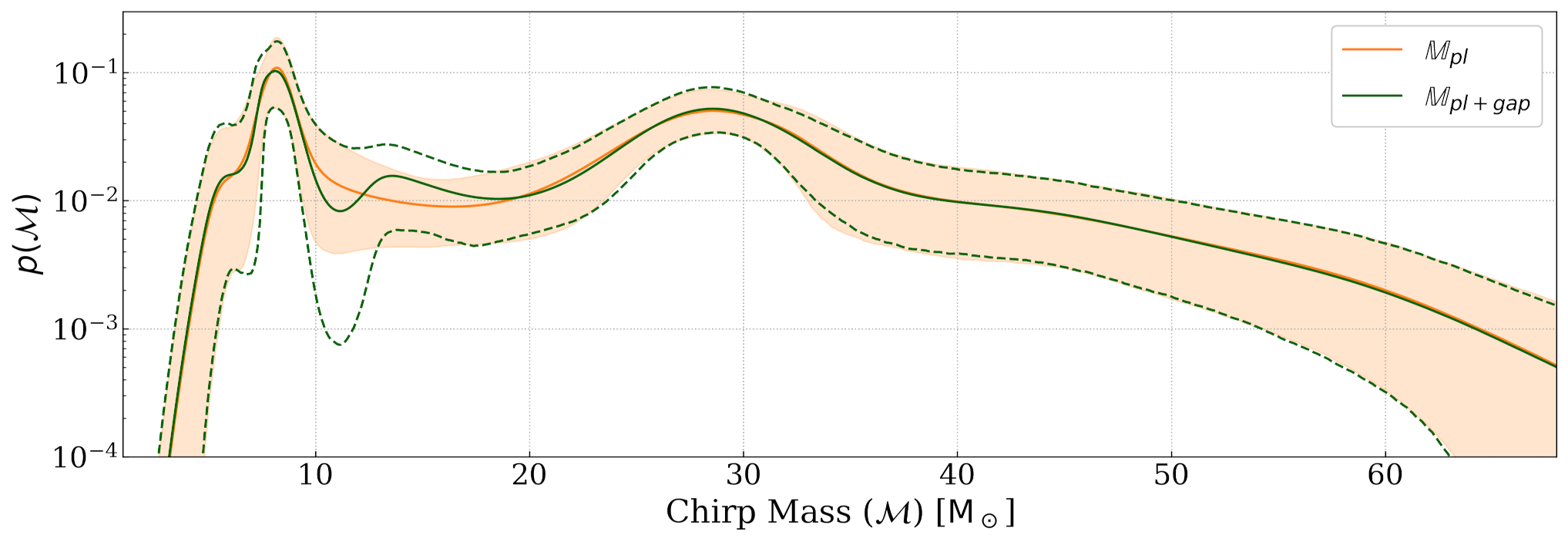}
    \caption{The chirp mass distribution inferred from the models $\mathbb{M}_{pl}$ and $\mathbb{M}_{pl+gap}$. The solid line represents the mean distribution, and the shaded region or dashed line indicates the 90\% credible interval. As expected, the orange curve does not exhibit features in the chirp mass range of 9.5--20 $ M_\odot$. It is used as the fiducial chirp mass distribution for creating simulated catalogs. The models infer almost identical distributions except in the range 9.5--20$M_\odot$.}
    \label{fig:pl_vs_plgap}.
\end{figure*}

\begin{figure*}
    \centering
    \includegraphics[width=\textwidth]{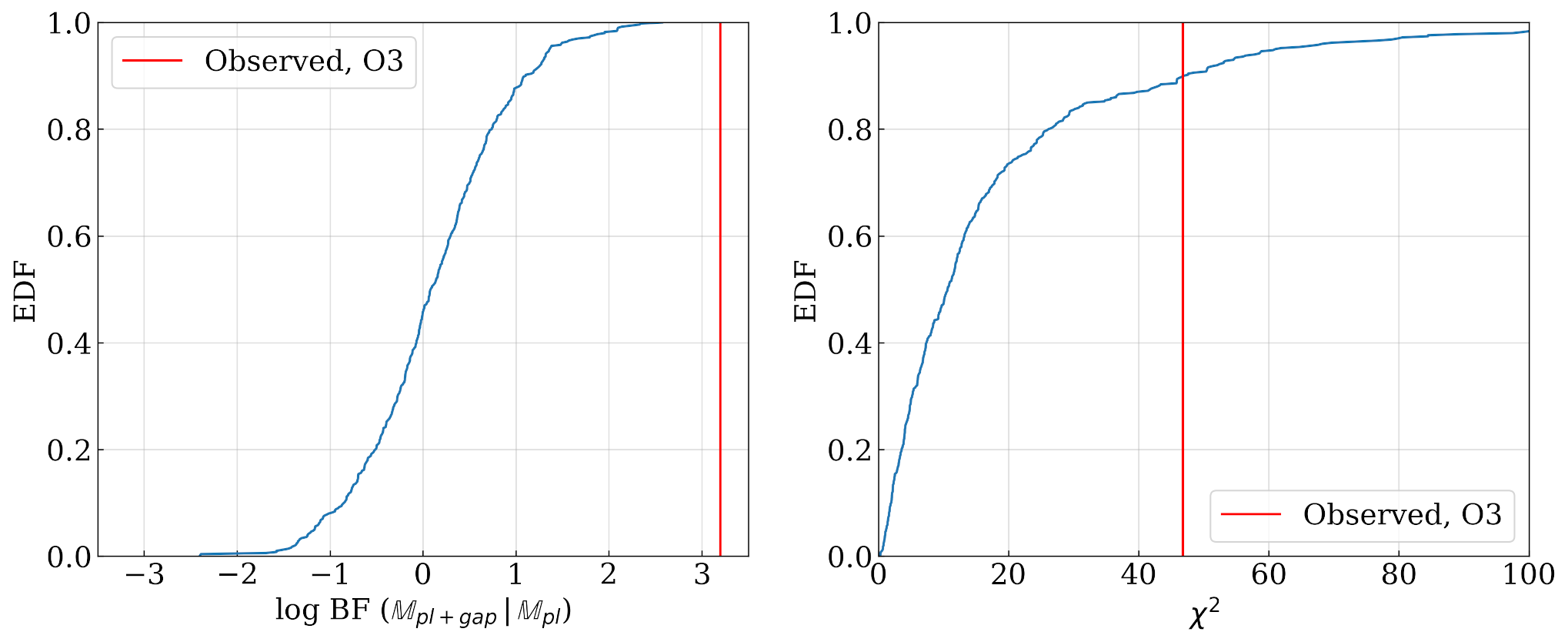}
    \caption{The log Bayes factor~(left) and $\chi^2$ values~(right) between models $\mathbb{M}_{pl+gap}$ and $\mathbb{M}_{pl}$ for the 500 simulated catalog. The red lines indicate the estimated values for the \ac{GW} observations.}
    \label{fig:analysis_pl_vs_plgap}
\end{figure*}

 Figure~\ref{fig:pl_vs_plgap} shows the inferred chirp mass distribution using models $\mathbb{M}_{pl}$ and $\mathbb{M}_{pl+gap}$. As expected, the distribution inferred from model $\mathbb{M}_{pl}$ does not show the presence of a local minimum or maximum in the chirp mass range 9.5--12$M_\odot$. However, the distribution inferred from model $\mathbb{M}_{pl+gap}$ shows a modulation. We calculate the Bayes factor and $\chi^2$ values for the \ac{GW} observations by using Equations~\ref{eq:BF} and \ref{eq:chisq}. To quantify confidence in this feature, we also calculate these values for the simulated catalogs. They are shown in Figure~\ref{fig:analysis_pl_vs_plgap} for the 500 catalogs we simulate. We find that none of the simulated catalog has a Bayes factor value larger than the one calculated for the observations. This suggests at least 99.5\% confidence in the presence of a gap. However, 10\% of the simulated catalogs have $\chi^2$ values larger than the one estimated for the observations, suggesting a smaller 90\% confidence in the presence of a gap. This is a conservative test, as we observe in the simulated catalogs that a large $\chi^2$ often arises due to an overall difference between the distributions inferred from the two models and not because of the presence of a local minimum in the distribution inferred by the model $\mathbb{M}_{pl+gap}$.

\subsection{Comparing $\mathbb{M}_{pl}$ and $\mathbb{M}$}

\begin{figure*}
    \centering
    \includegraphics[width=\textwidth]{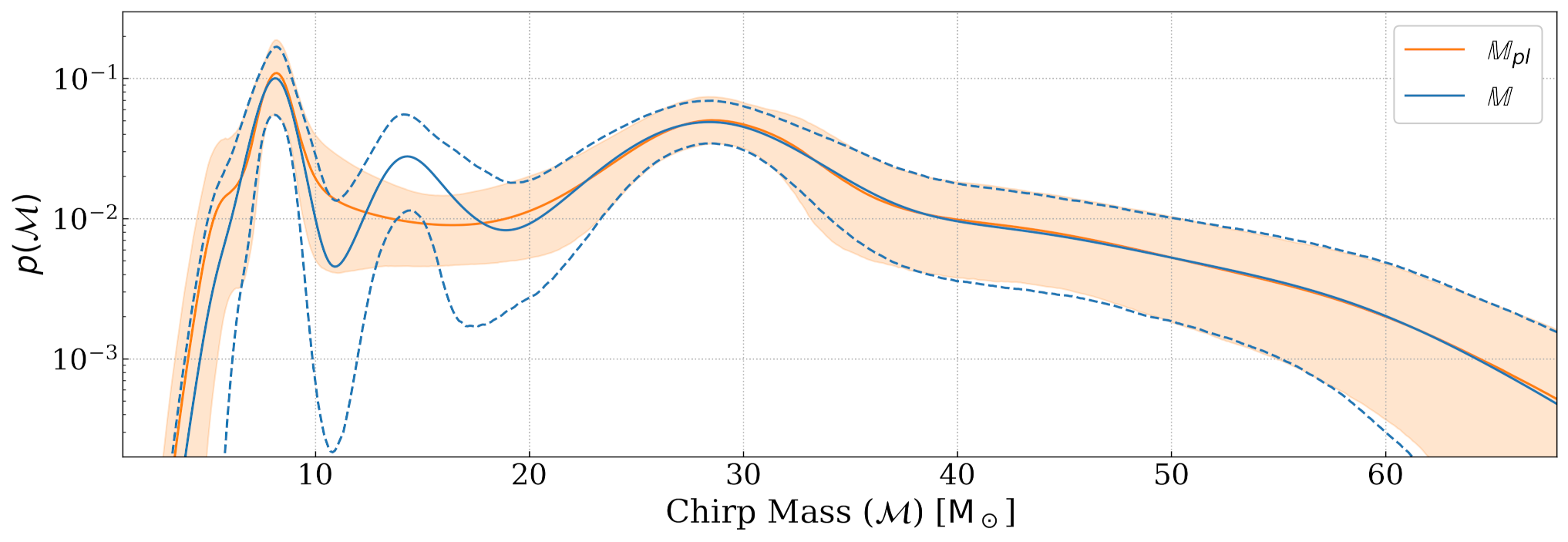}
    \caption{The chirp mass distribution inferred from the models $\mathbb{M}_{pl+gap}$ and $\mathbb{M}_{pl}$. The solid line represents the mean distribution, and the shaded region or dashed line indicates the 90\% credible interval. The models infer a similar distribution except in the range 9.5--20$M_\odot$.}
    \label{fig:mix_vs_pl}.
\end{figure*}

\begin{figure*}
    \centering
    \includegraphics[width=\textwidth]{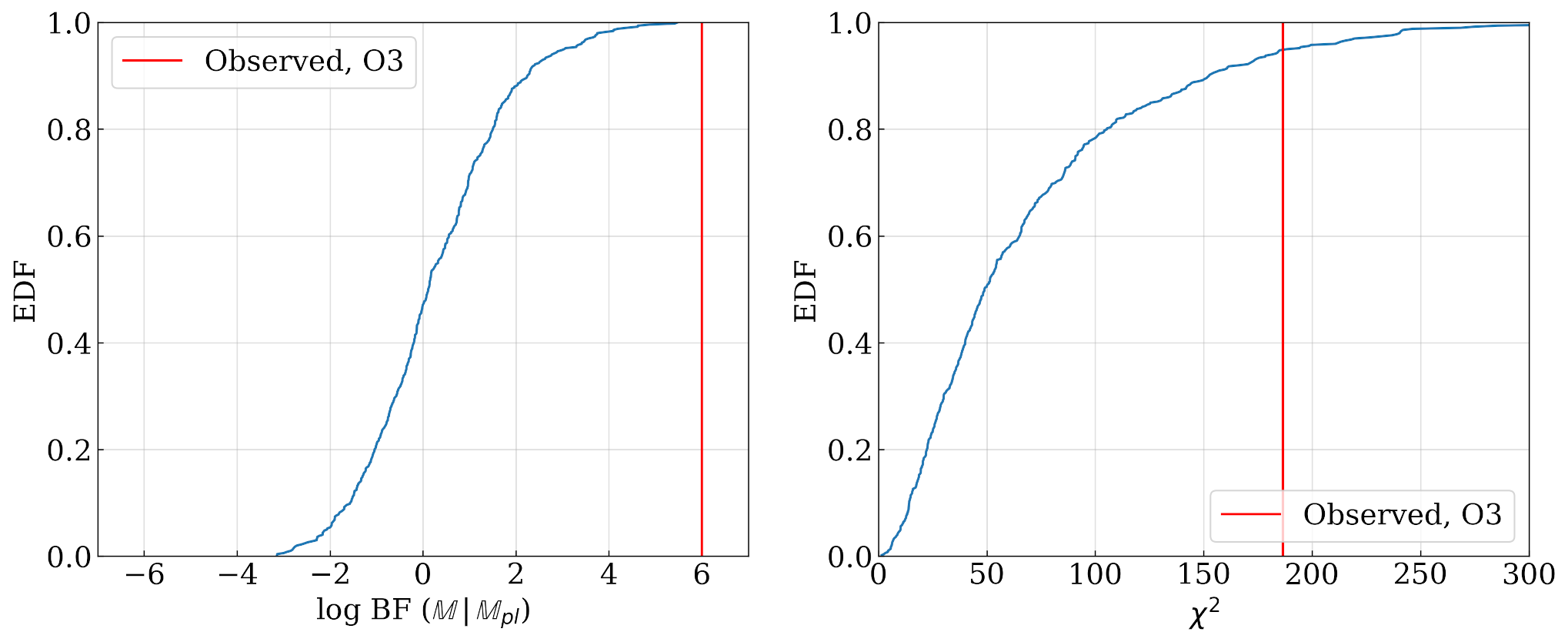}
    \caption{The log Bayes factor~(left) and $\chi^2$ values~(right) between models $\mathbb{M}_{pl+gap}$ and $\mathbb{M}_{pl}$ for the 500 simulated catalog. The red lines indicate the estimated values for the \ac{GW} observations.}
    \label{fig:analysis_mix_vs_pl}
\end{figure*}

Next, we repeat this procedure for $\mathbb{M}_{pl}$ and $\mathbb{M}$. Figure~\ref{fig:mix_vs_pl} shows the inferred chirp mass distribution using these models. Model $\mathbb{M}$ is more flexible in inferring a variety of distributions compared to the model $\mathbb{M}_{pl+gap}$. Its mean distribution shows a larger modulation than $\mathbb{M}_{pl+gap}$; however, the credible intervals are also larger. Figure~\ref{fig:analysis_mix_vs_pl} shows the Bayes factor and $\chi^2$ values between the two models for the simulated catalogs. We find that none of the simulated catalog has a Bayes factor value larger than the one calculated for the observations. This suggests at least 99.5\% confidence in the presence of a gap. However, 5\% catalogs have $\chi^2$ larger than the value for the \ac{GW} observation, suggesting a 95\% confidence in the presence of a gap. This is a conservative test, as $\chi^2$ can be large with or without the presence of a gap-like feature; i.e., a large $\chi^2$ only implies that two models have inferred distributions that differ. Unlike the previous section, which estimates the confidence in the presence of a gap characterised by suppression of density in a narrow chirp mass range, in this analysis, we are calculating the confidence in the feature inferred by the model $\mathbb{M}$. The mean distribution in Figure~\ref{fig:mix_vs_pl} suggests this feature may be a gap, or a gap followed by a peak.

\section{A Unique Correlation} \label{sec:uniqcorr}
\begin{figure*}
    \centering
    \includegraphics[width=0.47\textwidth]{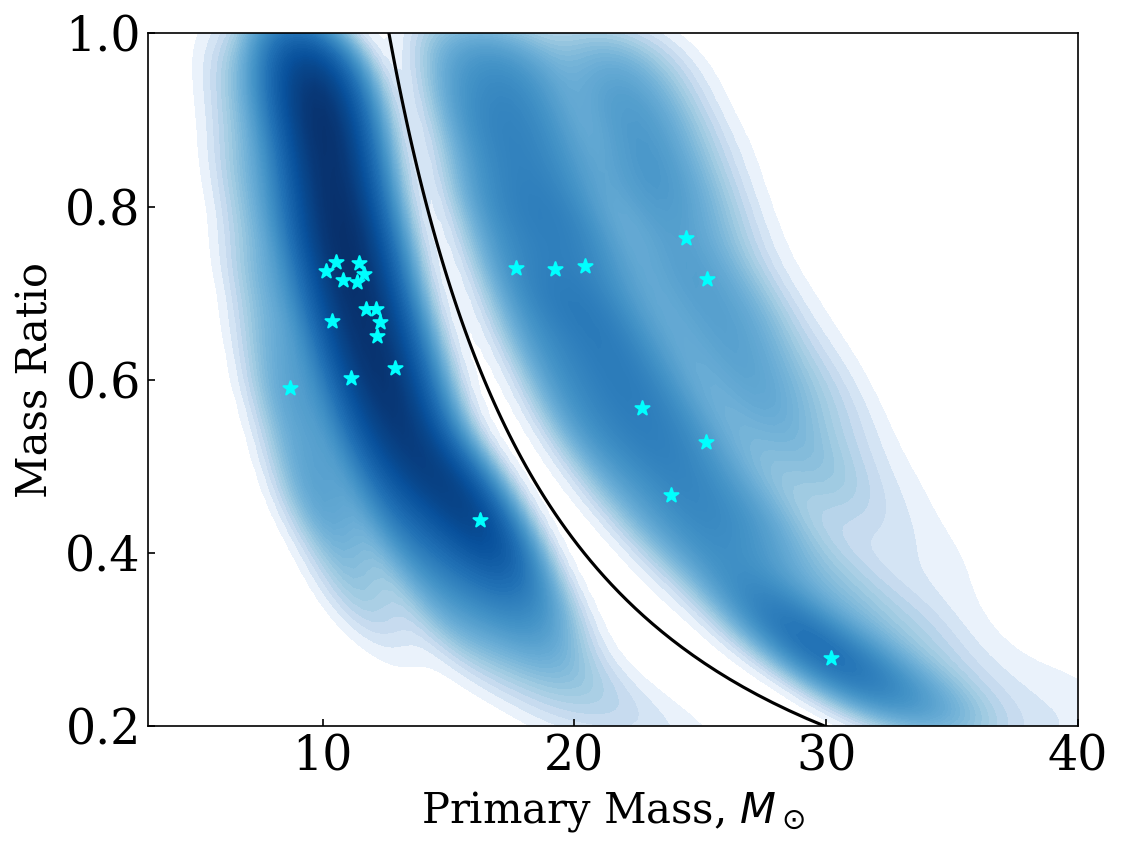}
    \includegraphics[width=0.52\textwidth]{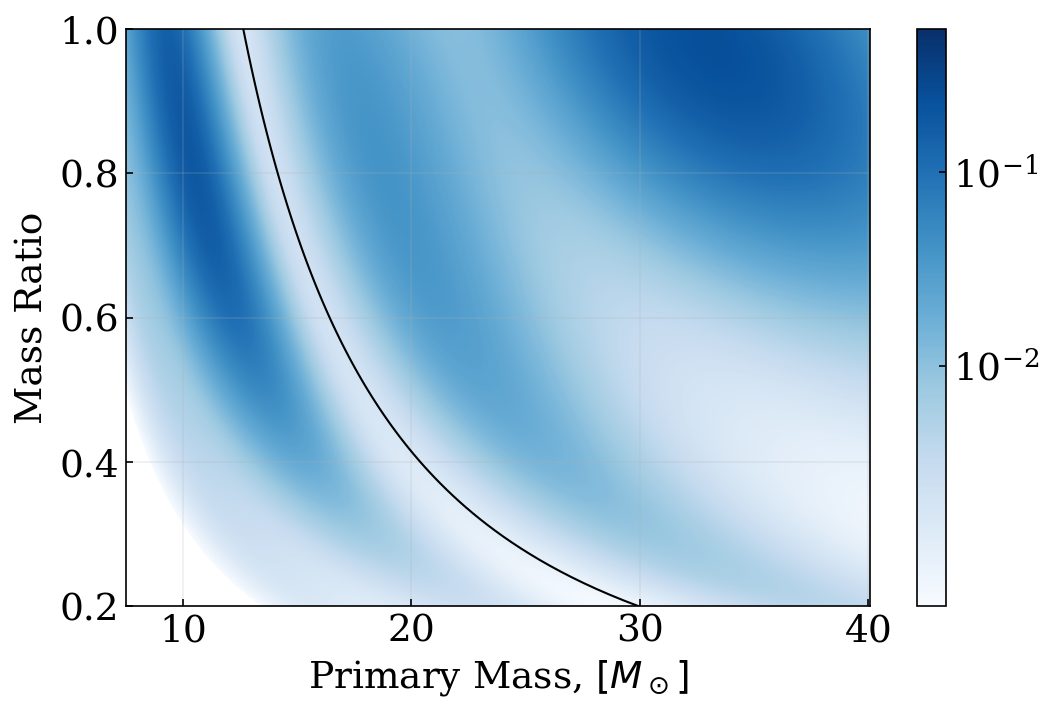}
    \caption{Left) The mass estimates of \acp{BBH} measured from the \acp{GW} stacked a 2D density plot. Only binaries measured with a mean chirp mass larger than 5$M_\odot$ and smaller than 20$M_\odot$ have been included. The stars are the median values measured for the individual \acp{BBH}. Right) The posterior predictive distribution obtained using the model $\mathbb{M}$. The black curve shows a constant $\mathcal{M} = 11.0M_\odot$ line.}
    \label{fig:m1vsm2}
\end{figure*}

Based on our tests, we can conclude that observations in GWTC-3 suggest a gap in the chirp mass range 9.5--12 $M_\odot$. In their analysis \cite{2024ApJ...975..253A} found no evidence for the presence of a gap in the component mass range 10--15$M_\odot$. They raise an important point about the pairing function: a gap in the chirp mass distribution, when combined with a mass-ratio distribution that follows a power-law with a spectral index of $\beta = 2$, results in a much shallower gap in the primary or secondary mass distribution~(Please see their Appendix~A). Therefore, the presence of a gap in the chirp mass distribution does not necessarily imply a gap in the component mass distribution. Chirp mass is the most accurately measured parameter and is expected to be the least biased. But chirp mass is not directly related to the physical processes responsible for the formation and merger of \acp{BBH}. For a gap in chirp mass to exist, the primary and secondary masses must combine uniquely. A larger primary mass should pair with a smaller secondary mass, and vice versa. Figure~\ref{fig:m1vsm2} shows all the mass measurements made from the \acp{GW} stacked in a 2D density plot. The plot includes only binaries with a mean chirp mass between 5$M_\odot$ and 20$M_\odot$. This plot qualitatively shows a unique combination of the two masses that results in increased clustering of chirp masses around specific values compared to other mass parameters. A quantitative investigation will require considering the relatively weak measurability of features in the component distribution; something we intend to conduct with a larger \ac{GW} catalog. We estimated confidence in the presence of a peak around 14$M_\odot$ in \cite{2024MNRAS.527..298T}. A discussion similar to this sub-section is also present there.

\section{Discussion} 
\label{sec:discuss}
In this letter, we investigate the observed gap in the chirp mass distribution. We estimate the presence of a gap with a confidence level greater than 99.5\% based on a Bayes factor comparison. However, when comparing models using a much more conservative $\chi^2$ statistic, we infer the presence of a gap with a confidence level greater than 95\%. Independent analyses have reported a lack of evidence in favour of a gap in the comparable mass range for the component mass distribution. We qualitatively demonstrate that this can be due to a unique correlation between the primary and secondary mass values, resulting in increased clustering of chirp mass values around specific values compared to other mass parameters. The number of \ac{GW} are expected to quadruple in the next few months. This roughly implies a factor-of-two reduction in the size of the confidence bands for the inferred distributions. There will be a sufficient number of observations to make a robust estimation of the confidence in the primary features of the binary black hole mass distribution. Chirp mass is the most accurately measured parameter and the least affected by biases; however, because it is a function of the component masses, the presence of peaks or gaps in the one-dimensional chirp mass distribution appears as over-- or under--dense tracks in the two-dimensional component mass distribution. Therefore, the presence of a strong feature in the chirp mass distribution requires a careful explanation.

\begin{acknowledgments}
Thanks to Sukant Bose for LIGO's Publications \& Presentations review and to Alberto Vecchio and Tom Dent for the helpful feedback. This work was conducted on Cardiff University's HAWK HPC, supported by STFC grants ST/I006285/1 and ST/V005618/1.

This material is based upon work supported by NSF's LIGO Laboratory which is a major facility fully funded by the National Science Foundation.
\end{acknowledgments}

%





\appendix

\section{Simulating Gravitational Wave Catalog}
\label{sims}
In this analysis, the network comprises advanced LIGO Livingston and Hanford detectors~\citep{2015CQGra..32g4001L}. We draw uniformly on right ascension, declination, coalescence phase, polarisation angle, cosine of the inclination angle and square of the luminosity distance. The coalescence time is fixed to zero. Simulated injections are generated using {\sc{IMRPhenomD}}~\citep{2016PhRvD..93d4006H,phenomd2}. This model is optimised for binary systems with spins aligned with the orbital angular momentum. It contains only the leading (2,2) harmonic; therefore, precession and higher harmonics have not been considered. Thus, we draw chirp mass, mass ratio and two aligned spin components. Any signal that crosses a \ac{SNR} of 9.0 is tagged as \emph{observed} and selected for parameter estimation. We do not keep track of parameter draws but ensure that the \emph{observed} signals follow the distribution listed in Table~\ref{tab:param_dist}. The parameters are estimated using the non-Markovian parameter sampler Varaha~\citep{2023PhRvD.108b3001T, 2024arXiv240516568T}. The parameter sampler uses a uniform prior on the chirp mass, mass ratio, aligned spin components and the square of luminosity distance.

\begin{deluxetable}{cll}[b!]
\label{tab:param_dist}
\tablecaption{Distribution of the \emph{observed}~(corresponding to signals that from cross a network \ac{SNR} of 9.0) mass and spin parameters of the simulated population.}
\tablehead{
& \colhead{Parameter} & \colhead{Distribution} \\
}
\startdata
Chirp Mass & $\mathcal{M}$ & $p(\mathcal{M}) \propto \mathcal{M}^{-1/2}, \quad \mathcal{M} \in [4.0, 65.0] $ \\
Mass Ratio & $q$ & $p(q) \propto q^{-5/2}, \quad q \in [0.2, 1]$ \\
Aligned spin of the heavier object & $s_{1z}$ & $p(s_{1z}) = \mathcal{N}(\mu=0, \sigma=0.15)$ \\
Aligned spin of the lighter object & $s_{2z}$ & $p(s_{2z}) = \mathcal{N}(\mu=0, \sigma=0.15)$ \\
\enddata
\end{deluxetable}

The network \ac{SNR} threshold used to tag simulated signals as \emph{observed} has been chosen carefully. Parameters for a lower \ac{SNR} signal are estimated with a larger error than those with a higher \ac{SNR}. A threshold of 9.0 has been chosen to ensure comparable measurement errors between simulated injections and \ac{GW} measurements. Moreover, parameter estimation uses the standard Bayesian analysis; thus, parameters retain a similar correlation as observed in real signals. We performed 2,500 parameter estimation runs with masses and spins distribution listed in Table~\ref{tab:param_dist}. We maintain the mass ratio and spin distribution intact for the investigations presented in this article, while drawing from the fiducial chirp mass distributions through importance sampling. 

\section{Models Used in Population Inference}
\label{models}

The Bayesian framework to infer the compact binary populations while accounting for measurement error and selection effect has been discussed extensively in the literature. Among early reports, Equation 13 in ~\citet{2019MNRAS.486.1086M} presents the relevant Bayes equation. Different approaches use different population models while keeping the Bayesian framework unchanged.

Vamana is a population model that uses a mixture of components. The inferred distributions are truncated between 0.05 and 1.0 for mass ratio and -0.9 and 0.9 for aligned spins. We assume that aligned spins are identically but independently distributed. We utilise 10 components for all inferences and employ identical settings when inferring astrophysical and simulated populations. The models used in our investigations are presented in Table~\ref{tab:Lambda}:

i) Model $\mathbb{M}_{\mathcal{M}}$: This model infers the joint distribution of chirp mass, mass ratio and aligned spin distributions. Components are multivariate normals in this model. The cross-terms for the covariance matrix between different parameters are set to zero, except for the covariance between the chirp mass and the mass ratio. This allows the mixture model to infer the variation of mass ratio with chirp mass flexibly. Thus, model $\mathbb{M}_{\mathcal{M}}$ essentially consists of components composed of a bivariate normal and a univariate normal distribution. This is described in Equation~\ref{eq:model1}.
\begin{equation}
p(\mathcal{M}, q, \chi_1, \chi_2|\Lambda) = \sum_{i=1}^N w_i \;\mathcal{N}(\mathcal{M}, q|\mu_i^{\mathcal{M}}, \sigma_i^{\mathcal{M}}, \mu_i^q, \sigma_i^q, C_i^{\mathcal{M}q})\phi(\chi_1|\mu^\chi_i, \sigma^\chi_i)\;\phi(\chi_2|\mu^\chi_i, \sigma^\chi_i),
\label{eq:model1}
\end{equation}
where $N$ is the number of components, $w$ are the weights for the components in the mixture, $C^{\mathcal{M} q}$ is the covariance between the chirp mass parameter and mass ratio, $\mu$ and $\sigma$ are the location and scale of the Gaussians. 
\\\\
\noindent
ii) Model $\mathbb{M}_{pl}$: This model restricts the location of Gaussians to enforce an inference that is featureless between the two prominent peaks observed in the mass distribution. The intermediate-mass range is inferred using one component composed of a power-law to infer the chirp mass distribution and two univariate normals to infer the aligned spin and mass ratio distributions.

\begin{multline}
p(\mathcal{M}, q, \chi_1, \chi_2|\Lambda) = \prod_{i=1}^N w_i\,f(\mathcal{M}, q)\,\phi(\chi_1|\mu^\chi_i, \sigma^\chi_i)\;\phi(\chi_2|\mu^\chi_i, \sigma^\chi_i),\\
f(\mathcal{M}, q)=\begin{cases}
			\mathcal{N}(\mathcal{M}, q|\mu_i^{\mathfrak{M}}, \sigma_i^{\mathfrak{M}}, \mu_i^q, \sigma_i^q, C_i^{\mathcal{M}q}), & i = 1\cdots N-1\\
            \mathcal{P}(\mathcal{M}|\alpha^\mathcal{M}, \mathcal{M}_\mathrm{min}, \mathcal{M}_\mathrm{max})\,\phi(q|\mu_N^q, \sigma_N^q), & i = N
		 \end{cases}
\label{eq:model3}
\end{multline}

\begin{figure*}
    \centering
    \includegraphics[width=0.98\textwidth]{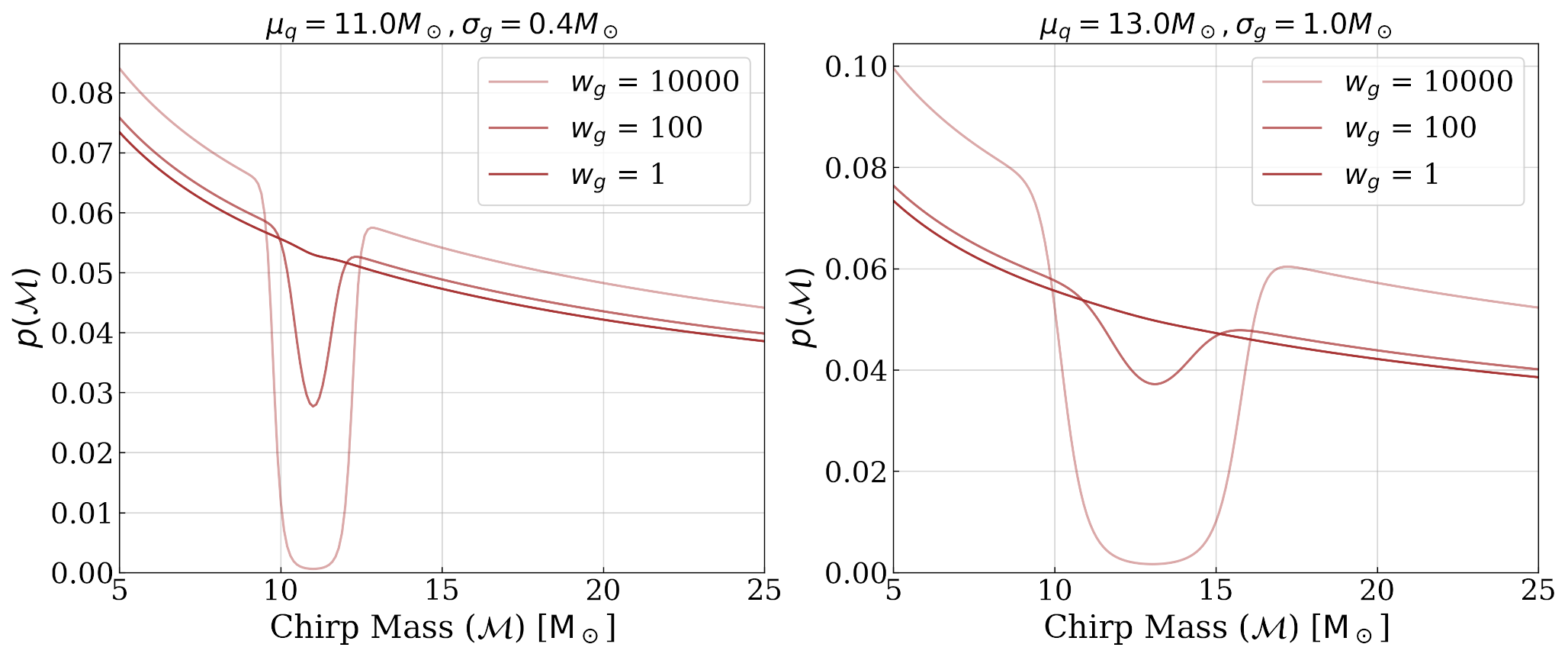}
    \caption{This figure plots the function defined in the Equation~\ref{eq:gap} for several values of hyperparameter. The power law truncates at 5.0$M_\odot$ and 25$M_\odot$ and has a spectral index of -0.4.}
    \label{fig:gapmorphology}
\end{figure*}

The model $\mathbb{M}_{pl+gap}$ is identical to $\mathbb{M}_{pl}$ but includes extra hyperparameters to suppress density in the narrow chirp mass range. The suppressed density is defined as,
\begin{equation}
    \label{eq:gap}
    p(\mathcal{M}) = \frac{p_{pl}(\mathcal{M})}{1 + w_g\,\phi(\mu_g, \sigma_g)/100},
\end{equation}
where $p_{pl}(\mathcal{M})$ is the power-law distribution and $\phi$ is the normal distribution with mean $\mu_g$ and scale $\sigma_g$. These hyperparameters control the location and width of the gap. Hyperparameter $w_g$ controls the depth of the gap. We use priors, $\mu_g \sim \mathcal{U}(9M_\odot, 20M_\odot)$, $\sigma_g \sim \mathcal{U}(0.2M_\odot, 1.0M_\odot)$, and $w_g \sim \text{Log-}\mathcal{U}(1, 10000)$. Figure~\ref{fig:gapmorphology} plots the distribution in Equation~\ref{eq:gap} for several values of hyperparameters.

The \ac{BBH} population is inferred using the observations with a false alarm rate of at most once per year. We only used observations reported by the LVK collaborations~\citep{2019PhRvX...9c1040A, 2021PhRvX..11b1053A, 2024PhRvD.109b2001A, 2023PhRvX..13d1039A}. Only observations with a mean chirp mass greater than 5$M_\odot$ are used. We excluded GW190814 and any observation consistent with a binary neutron star and neutron star-black hole binary from our investigations. The total number of observations chosen is 69. In addition, to reduce computational cost, we did not correct for selection effects and fixed the redshift evolution to be uniform in comoving coordinates. We don't expect this correction to introduce a feature in the target mass range and impact our conclusions. The binary parameters are estimated in the detector frame; we assume the Planck15 cosmology \citep{2016A&A...594A..13P} to change to the source frame quantities. 

\def\arraystretch{1.3}
\begin{table*}
	\centering
	\caption{This table lists the hyperparameters of the three models used to infer the \ac{BBH} population. The last column identifies the model where the hyperparameters are used. U stands for Uniform, and UL for Uniform-in-log.}

	\begin{tabular}{lcccc} 
		\hline
		$\Lambda$ & Description/Modeled Parameter & Prior & Range & Models\\
		\hline
		$w_i$ & Mixing weights, $w$ & Dirichlet($\bm{\alpha}$), $\alpha_{1\cdots N}= 1/N$ & 0--1 & $\mathbb{M}$, $\mathbb{M}_{pl}$\\
		$\mu_i^q$ & Gaussian's location, $q$& UL & .05--1.04 & $\mathbb{M}$, $\mathbb{M}_{pl}$\\
		$\sigma_i^q$ & Gaussian's scale, $q$ & U & 0.1\,--\,1.0\,/$\sqrt{N}$ & $\mathbb{M}$, $\mathbb{M}_{pl}$\\
$\mu^\chi_i$&Gaussian's location, $s_z$ &U/UL& $|\chi_i| < 0.4$\;/\;$0.4 < |\chi_i| < 0.9$& $\mathbb{M}$, $\mathbb{M}_{pl}$\\
  $\sigma^\chi_i$&Gaussian's scale, $s_z$ &U&0.1--1.3/\,$\sqrt{N}$& $\mathbb{M}$, $\mathbb{M}_{pl}$\\
		$\mu_i^{\mathcal{M}}$ & Gaussian's location, $\mathcal{M}$& U & 5$M_\odot$--60$M_\odot$ & $\mathbb{M}$\\
		$\mu_i^{\mathcal{M}}$ & Gaussian's location, $\mathcal{M}$& U & 5$M_\odot$--9$M_\odot$, 25$M_\odot$--60$M_\odot$ & $\mathbb{M}_{pl}$\\
		$\sigma_i^{\mathcal{M}}$ & Gaussian's scale, $\mathcal{M}$ & U & 0.05\,$\mu_i^{\mathcal{M}}$/$\sqrt{N}$\;--\;0.185\,$\mu_i^{\mathcal{M}}$/$\sqrt{N}$ & $\mathbb{M}$, $\mathbb{M}_{pl}$\\
  $C_i^{\mathcal{M}q}$ & Covariance, $\mathcal{M}$--$q$ & U & -0.5\,$\sigma_i^{\mathcal{M}}\,\sigma_i^q$\;--\; 0.5\,$\sigma_i^{\mathcal{M}}\,\sigma_i^q$& $\mathbb{M}$, $\mathbb{M}_{pl}$\\
  $\mathcal{M}_\mathrm{min}$ & Powerlaw lower edge, $\mathcal{M}$ & U & 6$M_\odot$--9.5$M_\odot$& $\mathbb{M}_{pl}$\\
  $\mathcal{M}_\mathrm{max}$ & Powerlaw higher edge, $\mathcal{M}$ & U & 28.5$M_\odot$--80$M_\odot$& $\mathbb{M}_{pl}$\\
  $\alpha^{\mathcal{M}}$ & Powerlaw slope, $\mathcal{M}$ & U & -3\,--\,3& $\mathbb{M}_{pl}$\\
  \hline
  \hline
	\end{tabular}
 \label{tab:Lambda}
\end{table*}
\def\arraystretch{1}

\newpage
\bibliography{bibliography}{}

@ARTICLE{2024MNRAS.527..298T,
       author = {{Tiwari}, Vaibhav},
        title = "{What's in a binary black hole's mass parameter?}",
      journal = {\mnras},
         year = 2024,
        month = jan,
       volume = {527},
       number = {1},
        pages = {298-306},
          doi = {10.1093/mnras/stad3155},
archivePrefix = {arXiv},
       eprint = {2304.03498},
 primaryClass = {astro-ph.HE},
       adsurl = {https://ui.adsabs.harvard.edu/abs/2024MNRAS.527..298T}
}

@ARTICLE{2023ApJ...955..107F,
       author = {{Farah}, Amanda M. and {Edelman}, Bruce and {Zevin}, Michael and {Fishbach}, Maya and {Mar{\'\i}a Ezquiaga}, Jose and {Farr}, Ben and {Holz}, Daniel E.},
        title = "{Things That Might Go Bump in the Night: Assessing Structure in the Binary Black Hole Mass Spectrum}",
      journal = {\apj},
         year = 2023,
        month = oct,
       volume = {955},
       number = {2},
          eid = {107},
        pages = {107},
          doi = {10.3847/1538-4357/aced02},
archivePrefix = {arXiv},
       eprint = {2301.00834},
 primaryClass = {astro-ph.HE},
       adsurl = {https://ui.adsabs.harvard.edu/abs/2023ApJ...955..107F}
}

@ARTICLE{2021CQGra..38o5007T,
       author = {{Tiwari}, Vaibhav},
        title = "{VAMANA: modeling binary black hole population with minimal assumptions}",
      journal = {Classical and Quantum Gravity},
         year = 2021,
        month = aug,
       volume = {38},
       number = {15},
          eid = {155007},
        pages = {155007},
          doi = {10.1088/1361-6382/ac0b54},
archivePrefix = {arXiv},
       eprint = {2006.15047},
 primaryClass = {astro-ph.HE},
       adsurl = {https://ui.adsabs.harvard.edu/abs/2021CQGra..38o5007T}
}

@ARTICLE{2023ApJ...946...16E,
       author = {{Edelman}, Bruce and {Farr}, Ben and {Doctor}, Zoheyr},
        title = "{Cover Your Basis: Comprehensive Data-driven Characterization of the Binary Black Hole Population}",
      journal = {\apj},
         year = 2023,
        month = mar,
       volume = {946},
       number = {1},
          eid = {16},
        pages = {16},
          doi = {10.3847/1538-4357/acb5ed},
archivePrefix = {arXiv},
       eprint = {2210.12834},
 primaryClass = {astro-ph.HE},
       adsurl = {https://ui.adsabs.harvard.edu/abs/2023ApJ...946...16E}
}

@ARTICLE{o3b_rnp,
author = {LVK},
        title = "{Population of Merging Compact Binaries Inferred Using Gravitational Waves through GWTC-3}",
      journal = {Physical Review X},
         year = 2023,
        month = jan,
       volume = {13},
       number = {1},
          eid = {011048},
        pages = {011048},
          doi = {10.1103/PhysRevX.13.011048},
archivePrefix = {arXiv},
       eprint = {2111.03634},
 primaryClass = {astro-ph.HE},
       adsurl = {https://ui.adsabs.harvard.edu/abs/2023PhRvX..13a1048A}
}

@ARTICLE{2024arXiv240403166R,
       author = {{Ray}, Anarya and {Maga{\~n}a Hernandez}, Ignacio and {Breivik}, Katelyn and {Creighton}, Jolien},
        title = "{Searching for binary black hole sub-populations in gravitational wave data using binned Gaussian processes}",
      journal = {arXiv e-prints},
         year = 2024,
        month = apr,
          eid = {arXiv:2404.03166},
        pages = {arXiv:2404.03166},
          doi = {10.48550/arXiv.2404.03166},
archivePrefix = {arXiv},
       eprint = {2404.03166},
 primaryClass = {astro-ph.HE},
       adsurl = {https://ui.adsabs.harvard.edu/abs/2024arXiv240403166R}
}

@ARTICLE{2021ApJ...913L..19T,
       author = {{Tiwari}, Vaibhav and {Fairhurst}, Stephen},
        title = "{The Emergence of Structure in the Binary Black Hole Mass Distribution}",
      journal = {\apjl},
         year = 2021,
        month = jun,
       volume = {913},
       number = {2},
          eid = {L19},
        pages = {L19},
          doi = {10.3847/2041-8213/abfbe7},
archivePrefix = {arXiv},
       eprint = {2011.04502},
 primaryClass = {astro-ph.HE},
       adsurl = {https://ui.adsabs.harvard.edu/abs/2021ApJ...913L..19T}
}

@ARTICLE{2021ApJ...917...33L,
       author = {{Li}, Yin-Jie and {Wang}, Yuan-Zhu and {Han}, Ming-Zhe and {Tang}, Shao-Peng and {Yuan}, Qiang and {Fan}, Yi-Zhong and {Wei}, Da-Ming},
        title = "{A Flexible Gaussian Process Reconstruction Method and the Mass Function of the Coalescing Binary Black Hole Systems}",
      journal = {\apj},
         year = 2021,
        month = aug,
       volume = {917},
       number = {1},
          eid = {33},
        pages = {33},
          doi = {10.3847/1538-4357/ac0971},
archivePrefix = {arXiv},
       eprint = {2104.02969},
 primaryClass = {astro-ph.HE},
       adsurl = {https://ui.adsabs.harvard.edu/abs/2021ApJ...917...33L}
}

@ARTICLE{2022ApJ...928..155T,
       author = {{Tiwari}, Vaibhav},
        title = "{Exploring Features in the Binary Black Hole Population}",
      journal = {\apj},
         year = 2022,
        month = apr,
       volume = {928},
       number = {2},
          eid = {155},
        pages = {155},
          doi = {10.3847/1538-4357/ac589a},
archivePrefix = {arXiv},
       eprint = {2111.13991},
 primaryClass = {astro-ph.HE},
       adsurl = {https://ui.adsabs.harvard.edu/abs/2022ApJ...928..155T}
}

@ARTICLE{2024PhRvL.133e1401L,
       author = {{Li}, Yin-Jie and {Wang}, Yuan-Zhu and {Tang}, Shao-Peng and {Fan}, Yi-Zhong},
        title = "{Resolving the Stellar-Collapse and Hierarchical-Merger Origins of the Coalescing Black Holes}",
      journal = {\prl},
         year = 2024,
        month = aug,
       volume = {133},
       number = {5},
          eid = {051401},
        pages = {051401},
          doi = {10.1103/PhysRevLett.133.051401},
archivePrefix = {arXiv},
       eprint = {2303.02973},
 primaryClass = {astro-ph.HE},
       adsurl = {https://ui.adsabs.harvard.edu/abs/2024PhRvL.133e1401L}
}

@ARTICLE{2023MNRAS.524.5844T,
       author = {{Toubiana}, A. and {Katz}, Michael L. and {Gair}, Jonathan R.},
        title = "{Is there an excess of black holes around 20 M{\ensuremath{\odot}}? Optimizing the complexity of population models with the use of reversible jump MCMC.}",
      journal = {\mnras},
         year = 2023,
        month = oct,
       volume = {524},
       number = {4},
        pages = {5844-5853},
          doi = {10.1093/mnras/stad2215},
archivePrefix = {arXiv},
       eprint = {2305.08909},
 primaryClass = {gr-qc},
       adsurl = {https://ui.adsabs.harvard.edu/abs/2023MNRAS.524.5844T}
}

@ARTICLE{2023arXiv230401288G,
       author = {{Godfrey}, Jaxen and {Edelman}, Bruce and {Farr}, Ben},
        title = "{Cosmic Cousins: Identification of a Subpopulation of Binary Black Holes Consistent with Isolated Binary Evolution}",
      journal = {arXiv e-prints},
         year = 2023,
        month = apr,
          eid = {arXiv:2304.01288},
        pages = {arXiv:2304.01288},
          doi = {10.48550/arXiv.2304.01288},
archivePrefix = {arXiv},
       eprint = {2304.01288},
 primaryClass = {astro-ph.HE},
       adsurl = {https://ui.adsabs.harvard.edu/abs/2023arXiv230401288G}
}

@article{2016PhRvD..93d4006H,
 adsurl = {https://ui.adsabs.harvard.edu/\\#abs/2016PhRvD..93d4006H},
 archiveprefix = {arXiv},
 author = {{Husa}, Sascha and {Khan}, Sebastian and {Hannam}, Mark and {P{\"u}rrer}, Michael and {Ohme}, Frank and {Forteza}, Xisco Jim{\'e}nez and {Boh{\'e}}, Alejandro},
 doi = {10.1103/PhysRevD.93.044006},
 eid = {044006},
 eprint = {1508.07250},
 journal = {\prd},
 month = {February},
 pages = {044006},
 primaryclass = {gr-qc},
 title = {{Frequency-domain gravitational waves from nonprecessing black-hole binaries. I. New numerical waveforms and anatomy of the signal}},
 url = {https://doi.org/10.1103/PhysRevD.93.044006},
 volume = {93},
 year = {2016}
}

@article{phenomd2,
        author = {Khan, S. and others},
        Title = {{Frequency-domain gravitational waves from nonprecessing black-hole binaries. II. A phenomenological model for the advanced detector era}},
        Url ={https://journals.aps.org/prd/abstract/10.1103/PhysRevD.93.044007},
        Journal = {Phys. Rev. D}, 
        Volume = {93}, 
        Number = {044006},
        Year = {2016}}

@article{2015CQGra..32g4001L,
 adsurl = {https://ui.adsabs.harvard.edu/\\#abs/2015CQGra..32g4001L},
 archiveprefix = {arXiv},
 author = {{Aasi}, J. and {Abbott}, B. P. and {Abbott}, R. and {Abbott}, T. and others},
 doi = {10.1088/0264-9381/32/7/074001},
 eid = {074001},
 eprint = {1411.4547},
 journal = {Classical and Quantum Gravity},
 month = {April},
 pages = {074001},
 primaryclass = {gr-qc},
 title = {{Advanced LIGO}},
 url = {https://doi.org/10.1088/0264-9381/32/7/074001},
 volume = {32},
 year = {2015}
}

@ARTICLE{2023PhRvD.108b3001T,
       author = {{Tiwari}, Vaibhav and {Hoy}, Charlie and {Fairhurst}, Stephen and {MacLeod}, Duncan},
        title = "{Fast non-Markovian sampler for estimating gravitational-wave posteriors}",
      journal = {\prd},
         year = 2023,
        month = jul,
       volume = {108},
       number = {2},
          eid = {023001},
        pages = {023001},
          doi = {10.1103/PhysRevD.108.023001},
archivePrefix = {arXiv},
       eprint = {2303.01463},
 primaryClass = {astro-ph.HE},
       adsurl = {https://ui.adsabs.harvard.edu/abs/2023PhRvD.108b3001T}
}

@ARTICLE{2024arXiv240516568T,
       author = {{Tiwari}, Vaibhav},
        title = "{Varaha: A promising sampler for obtaining gravitational wave posteriors}",
      journal = {arXiv e-prints},
         year = 2024,
        month = may,
          eid = {arXiv:2405.16568},
        pages = {arXiv:2405.16568},
          doi = {10.48550/arXiv.2405.16568},
archivePrefix = {arXiv},
       eprint = {2405.16568},
 primaryClass = {astro-ph.HE},
       adsurl = {https://ui.adsabs.harvard.edu/abs/2024arXiv240516568T}
}

@ARTICLE{2019PhRvX...9c1040A,
       author = {{LIGO Scientific Collaboration} and
         {Virgo Collaboration}},
        title = "{GWTC-1: A Gravitational-Wave Transient Catalog of Compact Binary Mergers Observed by LIGO and Virgo during the First and Second Observing Runs}",
      journal = {Physical Review X},
         year = 2019,
        month = jul,
       volume = {9},
       number = {3},
          eid = {031040},
        pages = {031040},
          doi = {10.1103/PhysRevX.9.031040},
archivePrefix = {arXiv},
       eprint = {1811.12907},
 primaryClass = {astro-ph.HE},
       adsurl = {https://ui.adsabs.harvard.edu/abs/2019PhRvX...9c1040A}
}

@ARTICLE{2021PhRvX..11b1053A,
       author = {{Abbott}, R. and {Abbott}, T.~D. and {Abraham}, S. and others},
        title = "{GWTC-2: Compact Binary Coalescences Observed by LIGO and Virgo during the First Half of the Third Observing Run}",
      journal = {Physical Review X},
         year = 2021,
        month = apr,
       volume = {11},
       number = {2},
          eid = {021053},
        pages = {021053},
          doi = {10.1103/PhysRevX.11.021053},
archivePrefix = {arXiv},
       eprint = {2010.14527},
 primaryClass = {gr-qc},
       adsurl = {https://ui.adsabs.harvard.edu/abs/2021PhRvX..11b1053A}
}

@ARTICLE{2023PhRvX..13d1039A,
       author = {{Abbott}, R. and {Abbott}, T.~D. and {Acernese}, F. and others},
        title = "{GWTC-3: Compact Binary Coalescences Observed by LIGO and Virgo during the Second Part of the Third Observing Run}",
      journal = {Physical Review X},
         year = 2023,
        month = oct,
       volume = {13},
       number = {4},
          eid = {041039},
        pages = {041039},
          doi = {10.1103/PhysRevX.13.041039},
archivePrefix = {arXiv},
       eprint = {2111.03606},
 primaryClass = {gr-qc},
       adsurl = {https://ui.adsabs.harvard.edu/abs/2023PhRvX..13d1039A}
}

@ARTICLE{2016A&A...594A..13P,
       author = {{Ade}, P.~A.~R. and {Aghanim}, N. and {Arnaud}, M. and others},
        title = "{Planck 2015 results. XIII. Cosmological parameters}",
      journal = {\aap},
         year = 2016,
        month = sep,
       volume = {594},
          eid = {A13},
        pages = {A13},
          doi = {10.1051/0004-6361/201525830},
archivePrefix = {arXiv},
       eprint = {1502.01589},
 primaryClass = {astro-ph.CO},
       adsurl = {https://ui.adsabs.harvard.edu/abs/2016A&A...594A..13P}
}

@article{Cutler:1994ys,
        author         = {Cutler, Curt and Flanagan, Eanna E.},
        title          = {{Gravitational waves from merging compact binaries: How
                        accurately can one extract the binary's parameters from
                        the inspiral wave form?}},
        journal        = {Phys. Rev. D},
        volume         = {49},
        year           = {1994},
        pages          = {2658-2697},
        doi            = {10.1103/PhysRevD.49.2658},
        eprint         = {9402014},
        archivePrefix  = {arXiv},
        primaryClass   = {gr-qc},
        reportNumber   = {GRP-369},
        Url = {https://doi.org/10.1103/PhysRevD.49.2658}
}

@ARTICLE{2021A&A...645A...5S,
       author = {{Schneider}, F.~R.~N. and {Podsiadlowski}, Ph. and {M{\"u}ller}, B.},
        title = "{Pre-supernova evolution, compact-object masses, and explosion properties of stripped binary stars}",
      journal = {\aap},
         year = 2021,
        month = jan,
       volume = {645},
          eid = {A5},
        pages = {A5},
          doi = {10.1051/0004-6361/202039219},
archivePrefix = {arXiv},
       eprint = {2008.08599},
 primaryClass = {astro-ph.SR},
       adsurl = {https://ui.adsabs.harvard.edu/abs/2021A&A...645A...5S}
}

@ARTICLE{2023ApJ...950L...9S,
       author = {{Schneider}, Fabian R.~N. and {Podsiadlowski}, Philipp and {Laplace}, Eva},
        title = "{Bimodal Black Hole Mass Distribution and Chirp Masses of Binary Black Hole Mergers}",
      journal = {\apjl},
         year = 2023,
        month = jun,
       volume = {950},
       number = {2},
          eid = {L9},
        pages = {L9},
          doi = {10.3847/2041-8213/acd77a},
archivePrefix = {arXiv},
       eprint = {2305.02380},
 primaryClass = {astro-ph.HE},
       adsurl = {https://ui.adsabs.harvard.edu/abs/2023ApJ...950L...9S}
}

@ARTICLE{2024ApJ...975..253A,
       author = {{Adamcewicz}, Christian and {Lasky}, Paul D. and {Thrane}, Eric and {Mandel}, Ilya},
        title = "{No Evidence for a Dip in the Binary Black Hole Mass Spectrum}",
      journal = {\apj},
         year = 2024,
        month = nov,
       volume = {975},
       number = {2},
          eid = {253},
        pages = {253},
          doi = {10.3847/1538-4357/ad7ea8},
archivePrefix = {arXiv},
       eprint = {2406.11111},
 primaryClass = {astro-ph.HE},
       adsurl = {https://ui.adsabs.harvard.edu/abs/2024ApJ...975..253A}
}

@ARTICLE{2025A&A...694A.186G,
       author = {{Galaudage}, Shanika and {Lamberts}, Astrid},
        title = "{Compactness peaks: An astrophysical interpretation of the mass distribution of merging binary black holes}",
      journal = {\aap},
         year = 2025,
        month = feb,
       volume = {694},
          eid = {A186},
        pages = {A186},
          doi = {10.1051/0004-6361/202451654},
archivePrefix = {arXiv},
       eprint = {2407.17561},
 primaryClass = {astro-ph.HE},
       adsurl = {https://ui.adsabs.harvard.edu/abs/2025A&A...694A.186G}
}

@ARTICLE{2025PhRvD.111b3013M,
       author = {{Mahapatra}, Parthapratim and {Chattopadhyay}, Debatri and {Gupta}, Anuradha and {Favata}, Marc and {Sathyaprakash}, B.~S. and {Arun}, K.~G.},
        title = "{Predictions of a simple parametric model of hierarchical black hole mergers}",
      journal = {\prd},
         year = 2025,
        month = jan,
       volume = {111},
       number = {2},
          eid = {023013},
        pages = {023013},
          doi = {10.1103/PhysRevD.111.023013},
archivePrefix = {arXiv},
       eprint = {2209.05766},
 primaryClass = {astro-ph.HE},
       adsurl = {https://ui.adsabs.harvard.edu/abs/2025PhRvD.111b3013M}
}

@ARTICLE{2015CQGra..32b4001A,
       author = {{Acernese}, F. and others},
        title = "{Advanced Virgo: a second-generation interferometric gravitational wave detector}",
      journal = {Classical and Quantum Gravity},
         year = 2015,
        month = jan,
       volume = {32},
       number = {2},
          eid = {024001},
        pages = {024001},
          doi = {10.1088/0264-9381/32/2/024001},
archivePrefix = {arXiv},
       eprint = {1408.3978},
 primaryClass = {gr-qc},
       adsurl = {https://ui.adsabs.harvard.edu/abs/2015CQGra..32b4001A}
}

@ARTICLE{2021PTEP.2021eA101A,
       author = {{Akutsu}, T. and others},
        title = "{Overview of KAGRA: Detector design and construction history}",
      journal = {Progress of Theoretical and Experimental Physics},
         year = 2021,
        month = may,
       volume = {2021},
       number = {5},
          eid = {05A101},
        pages = {05A101},
          doi = {10.1093/ptep/ptaa125},
archivePrefix = {arXiv},
       eprint = {2005.05574},
 primaryClass = {physics.ins-det},
       adsurl = {https://ui.adsabs.harvard.edu/abs/2021PTEP.2021eA101A}
}

@ARTICLE{2023A&A...676A..31D,
       author = {{Disberg}, P. and {Nelemans}, G.},
        title = "{Failed supernovae as a natural explanation for the binary black hole mass distribution}",
      journal = {\aap},
         year = 2023,
        month = aug,
       volume = {676},
          eid = {A31},
        pages = {A31},
          doi = {10.1051/0004-6361/202245693},
archivePrefix = {arXiv},
       eprint = {2306.14332},
 primaryClass = {astro-ph.HE},
       adsurl = {https://ui.adsabs.harvard.edu/abs/2023A&A...676A..31D}
}

@ARTICLE{2024PhRvD.109b2001A,
       author = {{Abbott}, R. and {Abbott}, T.~D. and {Acernese}, F. and others},
        title = "{GWTC-2.1: Deep extended catalog of compact binary coalescences observed by LIGO and Virgo during the first half of the third observing run}",
      journal = {\prd},
         year = 2024,
        month = jan,
       volume = {109},
       number = {2},
          eid = {022001},
        pages = {022001},
          doi = {10.1103/PhysRevD.109.022001},
archivePrefix = {arXiv},
       eprint = {2108.01045},
 primaryClass = {gr-qc},
       adsurl = {https://ui.adsabs.harvard.edu/abs/2024PhRvD.109b2001A}
}

@ARTICLE{2023ApJ...957...37R,
       author = {{Ray}, Anarya and {Hernandez}, Ignacio Maga{\~n}a and {Mohite}, Siddharth and {Creighton}, Jolien and {Kapadia}, Shasvath},
        title = "{Nonparametric Inference of the Population of Compact Binaries from Gravitational-wave Observations Using Binned Gaussian Processes}",
      journal = {\apj},
         year = 2023,
        month = nov,
       volume = {957},
       number = {1},
          eid = {37},
        pages = {37},
          doi = {10.3847/1538-4357/acf452},
archivePrefix = {arXiv},
       eprint = {2304.08046},
 primaryClass = {gr-qc},
       adsurl = {https://ui.adsabs.harvard.edu/abs/2023ApJ...957...37R}
}

@ARTICLE{2019MNRAS.486.1086M,
       author = {{Mandel}, Ilya and {Farr}, Will M. and {Gair}, Jonathan R.},
        title = "{Extracting distribution parameters from multiple uncertain observations with selection biases}",
      journal = {\mnras},
         year = 2019,
        month = jun,
       volume = {486},
       number = {1},
        pages = {1086-1093},
          doi = {10.1093/mnras/stz896},
archivePrefix = {arXiv},
       eprint = {1809.02063},
 primaryClass = {physics.data-an},
       adsurl = {https://ui.adsabs.harvard.edu/abs/2019MNRAS.486.1086M}
}

@ARTICLE{2025arXiv251007573W,
       author = {{Willcox}, Reinhold and {Schneider}, Fabian R.~N. and {Laplace}, Eva and {Podsiadlowski}, Philipp and {Maltsev}, Kiril and {Mandel}, Ilya and {Marchant}, Pablo and {Sana}, Hugues and {Li}, Tjonnie G.~F. and {Hertog}, Thomas},
        title = "{Good things always come in 3s: trimodality in the binary black-hole chirp-mass distribution supports bimodal black-hole formation}",
      journal = {arXiv e-prints},
         year = 2025,
        month = oct,
          eid = {arXiv:2510.07573},
        pages = {arXiv:2510.07573},
          doi = {10.48550/arXiv.2510.07573},
archivePrefix = {arXiv},
       eprint = {2510.07573},
 primaryClass = {astro-ph.SR},
       adsurl = {https://ui.adsabs.harvard.edu/abs/2025arXiv251007573W}
}
\bibliographystyle{aasjournal}



\end{document}